\documentclass[journal]{IEEEtran}

\usepackage{ifpdf}
\usepackage{cite}
\usepackage[pdftex]{graphicx}
\usepackage{amsmath}
\usepackage{array}
\usepackage{hyperref}
\usepackage[caption=false,font=footnotesize]{subfig}
\usepackage{fixltx2e}
\usepackage{float} 
\usepackage{booktabs}
\usepackage{stfloats}
\usepackage{url}
\begin{document}

\title{Fuzzy Information Seeded Region Growing for Automated Lesions After Stroke Segmentation in MR Brain Images}

\author{Mario~Pascual~González~\IEEEmembership{}}

\markboth{Biomedical Imaging Course 302; University of Málaga}%
{Shell \MakeLowercase{\textit{et al.}}: Bare Demo of IEEEtran.cls for IEEE Journals}

\maketitle

\begin{abstract}
In the realm of medical imaging, precise segmentation of stroke lesions from brain MRI images stands as a critical challenge with significant implications for patient diagnosis and treatment. Addressing this, our study introduces an innovative approach using a Fuzzy Information Seeded Region Growing (FISRG) algorithm. Designed to effectively delineate the complex and irregular boundaries of stroke lesions, the FISRG algorithm combines fuzzy logic with Seeded Region Growing (SRG) techniques, aiming to enhance segmentation accuracy.

The research involved three experiments to optimize the FISRG algorithm's performance, each focusing on different parameters to improve the accuracy of stroke lesion segmentation. The highest Dice score achieved in these experiments was 94.2\%, indicating a high degree of similarity between the algorithm's output and the expert-validated ground truth. Notably, the best average Dice score, amounting to 88.1\%, was recorded in the third experiment, highlighting the efficacy of the algorithm in consistently segmenting stroke lesions across various slices.

Our findings reveal the FISRG algorithm’s strengths in handling the heterogeneity of stroke lesions. However, challenges remain in areas of abrupt lesion topology changes and in distinguishing lesions from similar intensity brain regions. The results underscore the potential of the FISRG algorithm in contributing significantly to advancements in medical imaging analysis for stroke diagnosis and treatment.
\end{abstract}


\IEEEpeerreviewmaketitle

\section{Introduction}
\label{sec:introduction}

The precise detection and segmentation of lesions resulting from strokes in brain imaging data plays a pivotal role in managing stroke aftermath. These lesions, which can impact neural pathways crucial for vital functions and cognitive abilities, require early and accurate identification for the formulation of effective treatment plans. The significance of this task cannot be overstated, as the prompt and detailed recognition of post-stroke lesions bears directly on the patient's prognosis, treatment options, and ultimately, their chances of recovery and quality of life. This challenge transcends mere technicality, embodying a profound responsibility. The development and refinement of tools and methodologies for this purpose hold the promise not only to enhance the quality of medical care but also to significantly improve and extend the lives of those affected by stroke.

To navigate this challenge, the medical community has continually sought to refine diagnostic techniques. Within this evolution, Magnetic Resonance Imaging (MRI) has emerged as a superior modality since its inception in 1977. Offering detailed tissue contrast without the use of damaging radiation, MRI revolutionizes how we visualize and understand the intricate structures of the brain, thus becoming an essential tool in the neuro-oncologist's arsenal.

Building on the precision of MRI, Seeded Region Growing (SRG) introduced by Adams and Bischof in 1994, has been a foundational image segmentation technique  \cite{adams1994seeded}. The introduction of Fuzzy Information Seeded Region Growing (FISRG) marks a significant enhancement of SRG, embracing fuzzy logic to address the ambiguities in MR images  \cite{mehnert1997improved}. FISRG begins with an initial 'seed'—a pixel or a small region accurately labeled as stroke lesions. It then iteratively adds neighboring pixels to the growing region based on the similarity of intensity values, which are evaluated using fuzzy membership functions. This approach allows for a more nuanced and adaptable process that accounts for the imprecise boundaries of brain stroke lesions, thereby enhancing segmentation accuracy and facilitating better medical outcomes.

The primary goal of this research is to create and validate a FISRG-based segmentation algorithm enhanced with k-means clustering for seed selection and mathematical morphology for post-processing. This algorithm is tailored for segmentation of transversal T1-weighted MR brain images. These images reveal detailed anatomical structures such as white matter, gray matter, and the skull, as well as the pathology of interest—brain stroke lesions. 
\section{Related Works}
\par Chuin-Mu Wang \textit{et. al.} developed a technique using Fuzzy Information Seeded Region Growing (FISRG) to delineate various brain structures in MRI scans, which has significantly influenced the direction of this research \cite{WangFuzzy}. In their innovative approach, they employ a fuzzy logic predicate that considers multiple factors: the standard deviation across the entire image, the standard deviation within the region presently undergoing segmentation, and the Gaussian distribution of pixel intensity for similarity computations. This method enables the pixel-by-pixel comparison to the growing region with a high degree of precision. Furthermore, they refine their methodology with a post-processing step, which consolidates similar regions based on defined criteria, focusing on three main regions for potential merging.

\par In the current study, the fuzzy predicate is tailored specifically to the Gaussian distribution of pixel intensity to ascertain the similarity for segmentation purposes, diverging from Chuin-Mu Wang's method by omitting the variable region merging techniques. This decision stems from the hypothesis that stroke lesions may exist as singular or non-fragmented entities, thus necessitating a focused and streamlined approach. Additionally, both preprocessing and postprocessing steps were incorporated into this work, marking a distinct methodological departure from Wang's framework. These alterations reflect a strategic simplification and adaptation of the FISRG technique to suit the specific objectives and hypotheses under investigation in this research while preserving an accurate result. 
\section{Materials}

\subsection*{Hardware Configuration}

The computational framework for the segmentation experiments was established using a custom-designed workstation, which was assembled to meet the high computational demands of the project. The core of the system included a Gigabyte GeForce RTX 3060 WINDFORCE OC graphics card, which boasts 12GB GDDR6 memory for intricate image processing tasks. Data storage and access were facilitated by dual WD BLACK SN770 1TB SSDs, which take advantage of the high-speed PCIe Gen4 NVMe interface. The system's processing power was further augmented with 32GB of Corsair Vengeance DDR5 RAM with a 6000MHz clock rate, and an Intel Core i7-13700KF processor that operates at a base clock of 3.4 GHz, ensuring swift and efficient data handling for the extensive computational workload.

\subsection*{Software Environment}

Our experiments were supported by the Ubuntu 22.04 LTS operating system, chosen for its long-term support and stability. The coding and development phase was conducted using the Pycharm Professional IDE, which offers a sophisticated environment for Python coding, complemented by Google's robust cloud services.

\subsection*{Dataset}

The dataset employed for the segmentation algorithm was the Anatomical Tracings of Lesions After Stroke (ATLAS) R2.0, which includes a considerable number of patient MRIs. ATLAS v2.0 presents a comprehensive collection, consisting of 955 T1-weighted MRI scans, segmented into 655 training images with corresponding manually-segmented lesion masks and 300 test images. 
The data used belonged only to one patient \cite{atlas}. 

\subsection*{Programming Languages and Libraries}

The segmentation algorithm development was performed using Python version 3.9, leveraging its extensive repository of libraries for scientific computing and image analysis. NumPy was utilized for high-level mathematical functions and array operations, while Matplotlib was instrumental for data visualization. Data structures and analysis were handled using Pandas, whereas NiBabel was essential for neuroimaging file manipulation. The OpenCV library offered an array of computer vision functions necessary for image processing tasks. Additionally, scikit-image provided tools for advanced image processing capabilities, supporting the extensive computational requirements of the study.

\begin{figure*}[b]
\centering
\includegraphics[width=.87\textwidth]{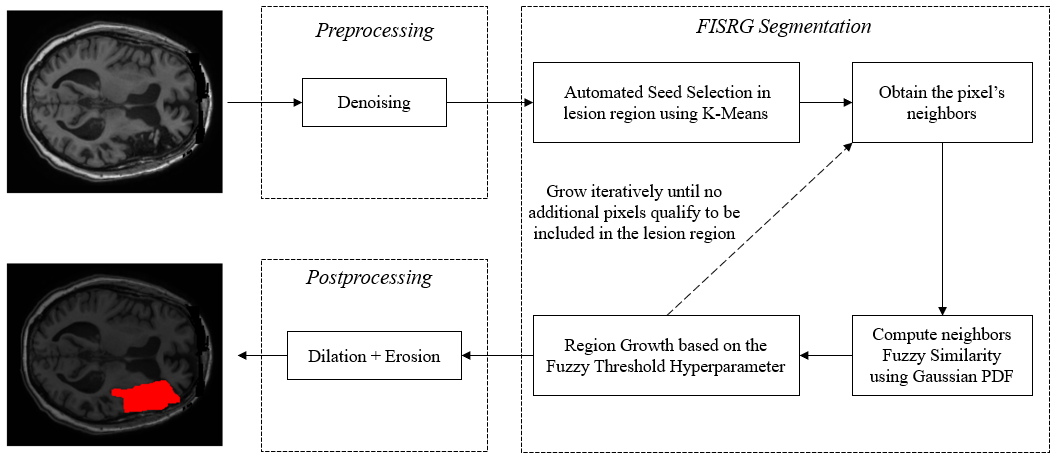}
\caption{Workflow of the Fuzzy Information Seeded Region Growing (FISRG) Segmentation Process applied in this research: The diagram illustrates the sequential stages in the FISRG methodology for MRI brain stroke lesions segmentation, starting with preprocessing through denoising, followed by FISRG segmentation including seed selection and region growing, and concluding with postprocessing involving morphological operations.}
\label{fig:diagramworkflow}
\end{figure*}

\section{Methodology}

The methodology adopted in this study is visualized in Fig. \ref{fig:diagramworkflow}. The process commences with a preprocessing stage where Gaussian denoising is applied to the MRI images to reduce noise and improve homogeneity within the stroke lesions region. Following this, the FISRG segmentation phase begins with the automated selection of seed points using K-Means clustering within the stroke lesions region, which then serves as a basis for region growing. This growth is driven by the computation of fuzzy similarity using a Gaussian PDF, iteratively incorporating pixels until the stroke lesions region is fully delineated. To refine the segmented regions, a postprocessing step of morphological closing, composed of dilation followed by erosion, is applied to smooth the segmentation results and enhance the accuracy of the stroke lesions boundaries. 

\subsection{Preprocessing}
In the preprocessing phase of the methodology, a Gaussian denoising technique was applied to the MRI images to enhance the quality of the data prior to segmentation. This step is crucial in mitigating the noise inherently present in MRI scans, which can obscure the subtle intensity variations indicative of pathological tissue. The Gaussian denoising process was implemented using a Gaussian filter, defined by the two-dimensional Gaussian function, \(G(x, y)\), shown in equation \ref{eq:GaussianFunction} that smooths the image by averaging pixel values, effectively reducing noise while preserving essential structural details. The application of this filter results in a more consistent intensity distribution within the stroke lesions region, facilitating the identification of homogeneous seed points for the FISRG algorithm. By smoothing the image data, the denoising step ensures that the subsequent region growing is not adversely affected by spurious noise, thereby enhancing the fidelity and reliability of the stroke lesions segmentation process. 

\begin{equation}
    G(x, y) = \frac{1}{2\pi\sigma^2} e^{-\frac{x^2 + y^2}{2\sigma^2}}
\label{eq:GaussianFunction}
\end{equation}

\begin{figure}[!t]
\centering
\includegraphics[width=.487\textwidth]{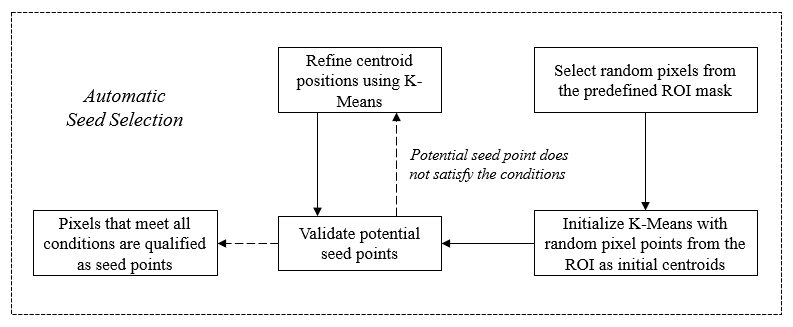}
\caption{Automatic Seed Selection Process in FISRG: This flowchart outlines the steps involved in the automatic selection of seed points for the FISRG algorithm. The process initiates with the selection of random pixels from the predefined ROI mask and employs K-Means clustering to refine centroid positions. These centroids are then validated against specific conditions to determine their suitability as seed points, ensuring that only those pixels that meet all criteria are qualified for subsequent segmentation stages.}
\label{fig:seedworkflow}
\end{figure}

\subsection{Seed Selection}
An integral component of the FISRG segmentation framework is the automatic identification of seed points from which the region growing process initiates. Fig. \ref{fig:seedworkflow} provides a comprehensive visual representation of this workflow, detailing the steps involved in the selection and validation of seed points within the Region of Interest (ROI).

The process involves identifying seed points that meet stringent criteria to serve as the foundation for the region growing algorithm. These seed points are selected to ensure homogeneity by evaluating the local intensity and variance, and they are spaced apart to maintain distinctiveness between the points.

The placement of seeds leverages the KMeans clustering algorithm, which segments the ROI into clusters aiming to reduce intra-cluster variance and enhance inter-cluster distinction. The algorithm's centroids represent the potential seed points. Yet, these centroids undergo a validation process to ascertain their suitability based on the predefined conditions of homogeneity and separation. Centroids meeting these conditions are then adopted as seeds for the ensuing region growing process, integral to the FISRG segmentation approach.

\subsection{Region Growth}

Upon establishing the initial seeds, the algorithm advances to the region growing phase, where each seed point catalyzes the expansion of the stroke lesions region. Employing an 8-connectivity model, the algorithm explores potential new region members among the immediate neighbors of each pixel, encompassing direct adjacencies as well as diagonal connections. 

Incorporating a pixel into the developing stroke lesions region hinges on a calculated fuzzy similarity measure derived from a Gaussian probability density function (PDF), \(S(x)\), shown in \ref{eq:GaussianPDF}. This measure quantifies the likelihood that a pixel's intensity aligns with the statistical profile of the stroke lesions—defined by the regional mean and standard deviation of intensities. Only those pixels whose fuzzy similarity scores surpass a specific threshold are assimilated into the region, confirming their concordance with the stroke lesion's established intensity pattern.

The fuzzy similarity measure is dynamically recalibrated as new pixels are added, with the algorithm adjusting the mean and standard deviation of the region's intensities in real-time. This adaptive updating is critical, as it ensures the growing region remains a true representation of the stroke lesions, even as it encounters heterogeneous tissue densities and complex patterns of growth. 

\begin{equation}
    S(x) = \frac{1}{\sqrt{2\pi\sigma^2}} e^{-\frac{(x - \mu)^2}{2\sigma^2}}
\label{eq:GaussianPDF}
\end{equation}

Finally, the FISRG algorithm concludes with the aggregation of individual regions grown from each seed point. The union of these regions forms a composite segmentation mask, representing the complete stroke lesions boundary as identified across all seeds.

\subsection{Postprocessing}
In the post-processing stage of the research, mathematical morphological operations—specifically, dilation followed by erosion—were applied to the segmented regions derived from the FISRG algorithm. We did not apply a 'closing' operation since the dilation kernel was bigger than the erosion kernel, this was due to the necessity of closing big gaps in the segmentation mask but with few necessity of erasing the contour. 

Dilation serves to connect disjointed regions and fill small holes within the segmented stroke lesions, effectively smoothing the boundaries and consolidating any fragmented areas. This operation is particularly beneficial for addressing the issue of segmentation underestimation, where small parts of the stroke lesions may be disconnected due to the variability in pixel intensity or noise remnants that were not entirely eliminated during preprocessing. 

Subsequent to dilation, erosion is applied, which aims to remove the thin layers of pixels added around the boundaries of the stroke lesions during dilation, restoring the region to its original size and shape. The combination of these two morphological operations helps in maintaining the structural integrity of the stroke lesions region, ensuring that the final segmentation is both robust and representative of the true pathology. 

\subsection{Hyperparameter fine-tuning}
The evaluation of the FISRG algorithm's performance in segmenting MRI images hinges on its alignment with expert-validated ground truth masks. To quantify this alignment, the Dice coefficient—a statistical tool that measures the similarity between two sets—was utilized. This coefficient is shown in equation \ref{eq:dice}, where \(X\) denotes the set of pixels in the ground truth mask, and \(Y\) represents the set of pixels in the predicted output mask, ranges from 0 to 1, where a DSC of 1 indicates perfect agreement between the predicted segmentation and the ground truth.

\begin{equation}
    DSC = \frac{2 |X \cap Y|}{|X| + |Y|}
\label{eq:dice}
\end{equation}

Parameter optimization was a methodical process that commenced with identifying variables influential to pixel intensity, such as the standard deviation (\(\sigma\)) value in Gaussian denoising, which directly impacts image smoothness. In addition to denoising parameters, the optimization focused on the algorithm's intrinsic hyperparameters: the number of seeds and the fuzzy threshold value. The number of seeds was adjusted to accommodate the heterogeneous nature of the stroke lesions, while the fuzzy threshold was critically evaluated for its role in governing the inclusivity of pixel addition to the growing stroke lesions region.

Furthermore, the postprocessing step, particularly the dilation operation, was optimized to address segmentation granularity. The kernel size in the dilation phase was fine-tuned to ensure that segmented regions did not retain holes, leading to a more coherent stroke lesions representation. Initially, parameter tuning employed a brute force approach; however, following preliminary data analysis, the optimization was refined to specified boundaries, enhancing efficiency and reducing computational overhead.

The paramount objective of this research was the accuracy of the segmentation, rather than computational speed, given that the immediate application is not in real-time clinical settings. Therefore, the optimization process was tailored to prioritize segmentation precision over processing time, ensuring the resulting algorithm can robustly handle the diversity presented in stroke lesions morphology and intensity profiles.

\begin{figure*}[!t]
\centering
\includegraphics[width=.9\textwidth]{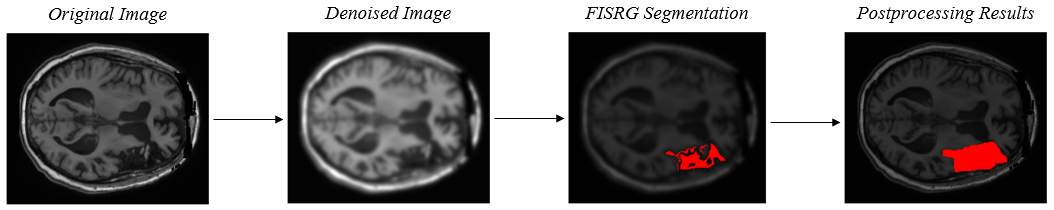}
\caption{MRI Image Segmentation Evaluation: The progression from the original MRI image through the various stages of the FISRG segmentation process. The sequence shows the original image, the denoised image after Gaussian smoothing, the initial FISRG segmentation result highlighting the stroke lesions area in red, and the final postprocessed image where morphological operations have refined the segmentation.}
\label{fig:workflowmethodology}
\end{figure*}

The evaluation of the segmentation process is depicted in Fig. \ref{fig:workflowmethodology}, which illustrates the transformation of a sample MRI image through the stages of the FISRG algorithm after optimization of the parameters. 

\subsection{Other Approaches Taken}
In the pursuit of optimizing the Dice coefficient and refining the segmentation accuracy of our FISRG algorithm, several alternative approaches were explored. These methodologies, while not adopted in the final implementation, contributed valuable insights into the segmentation process and the behavior of the algorithm under varying conditions. A comprehensive discussion of these ancillary methods, along with the rationale behind their trial and the specific outcomes of each, is provided in Appendix A. This detour through alternative strategies underscores the iterative nature of research and the exhaustive efforts undertaken to ensure the robustness of the segmentation results presented herein.
\section{Results}

In the evaluation of the FISRG algorithm, a series of three experiments were conducted, each utilizing the same dataset to ensure consistency across trials. The initial experiment focused on fine-tuning the fuzzy threshold and the number of seeds. The second experiment extended the parameter optimization to include the standard deviation (\(\sigma\)) of the Gaussian denoising kernel, in addition to the previously mentioned parameters. The final experiment further expanded the scope of optimization to encompass the dilation kernel size used in postprocessing, alongside the fuzzy threshold, number of seeds, and \(\sigma\).

Throughout these experiments, the optimal parameter values were identified for each MRI slice by comparing the algorithm's output against the ground truth mask and calculating the corresponding Dice coefficient. Additionally, the computational time for each iteration was meticulously recorded. This section will present the outcomes of these experiments, illustrating the algorithm's performance in terms of segmentation accuracy. The subsequent section will delve into a comprehensive discussion on the balance between computational efficiency and the distribution of the data, elucidating the trade-offs encountered during the study.

For each experiment depicted in this section, two primary data representations will be provided: graphical and tabular.

Firstly, a graph will be plotted for each experiment, which will illustrate the variation of the Dice coefficient across the MRI slices. This graphical representation will include a corresponding curve that delineates the real percentage of stroke lesions tissue present per slice. The inclusion of stroke lesions tissue distribution is intended to augment the analysis, providing a nuanced understanding of the segmentation performance in relation to stroke lesions prevalence within each slice.

Secondly, a comprehensive table will enumerate the statistical characteristics of the parameters optimized in each experiment, specifically detailing the fine-tuned fuzzy threshold, the achieved Dice coefficient, and the number of seeds utilized. This tabulation will serve to succinctly encapsulate the quantitative results of the fine-tuning process, facilitating a clear and concise comparison across the different experimental iterations. 

\subsection*{Experiment 1}
Experiment one established the baseline for the parameter optimization of the FISRG algorithm. In this initial phase, the primary focus was on the fine-tuning of the fuzzy threshold and the number of seed points used in the segmentation process. 

This experiment iterated through a range of values for both parameters, systematically searching for the combination that yielded the highest Dice coefficient, indicating the greatest similarity between the algorithm's segmentation and the ground truth. The results from this experiment laid the groundwork for subsequent experiments, providing insights into the delicate interplay between segmentation precision and algorithm efficiency.

\begin{figure}[H]
\centering
\includegraphics[width=.487\textwidth]{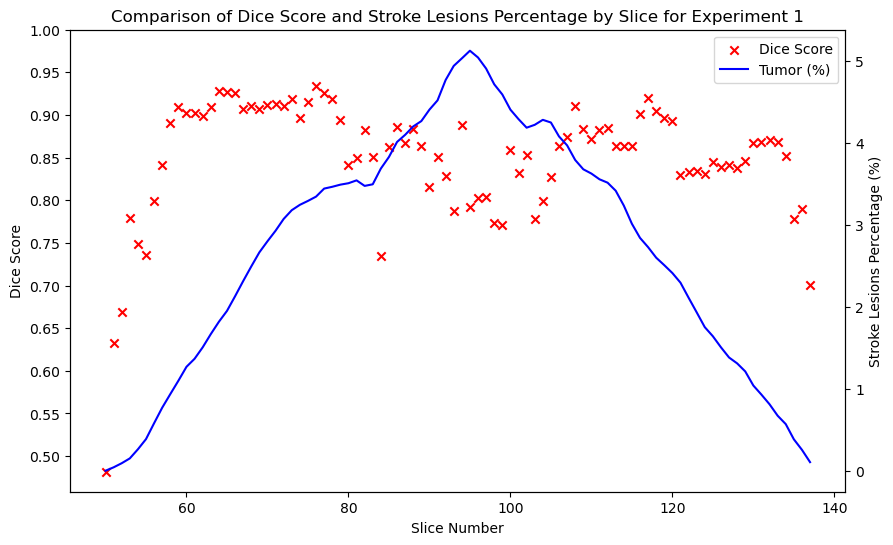}
\caption{Dice Score and stroke lesions Percentage per Slice in Experiment 1: The graph presents a comparison between the Dice score for each MRI slice and the corresponding percentage of stroke lesions tissue within that slice during the first experiment.}
\label{fig:exp1figure}
\end{figure}

\begin{table}[H]
\centering
\caption{Statistical Summary of Parameters for Experiment 1}
\label{tab:exp1data}
\begin{tabular}{@{}lccc@{}}
\toprule
 & \textbf{Fuzzy Threshold} & \textbf{Seeds} & \textbf{Dice Score} \\ 
\midrule
mean & 0.311 & 8.875 & 0.849 \\
std  & 0.239 & 2.486 & 0.071 \\
min  & 0.112 & 4.000 & 0.481 \\
max  & 1.000 & 11.000 & 0.934 \\
\bottomrule
\end{tabular}
\end{table}

\subsection*{Experiment 2}
Building upon the foundational work of experiment one, the second experiment extended the parameter optimization process to include the standard deviation (\(\sigma\)) of the Gaussian denoising kernel, in addition to the fuzzy threshold and the number of seeds. This experiment aimed to refine the denoising process, which is instrumental in enhancing the quality of the image prior to segmentation, and to assess the impact of these adjustments on the segmentation accuracy as quantified by the Dice coefficient.

The parameter \(\sigma\) plays a pivotal role in the Gaussian denoising stage, dictating the extent to which the image is smoothed and, consequently, the clarity of the stroke lesions boundaries for the following segmentation steps. Iterative fine-tuning of this parameter, along with the fuzzy threshold and the number of seeds, was conducted to identify the optimal combination that would yield the highest Dice score when compared to the ground truth mask for each slice.

\begin{figure}[H]
\centering
\includegraphics[width=.487\textwidth]{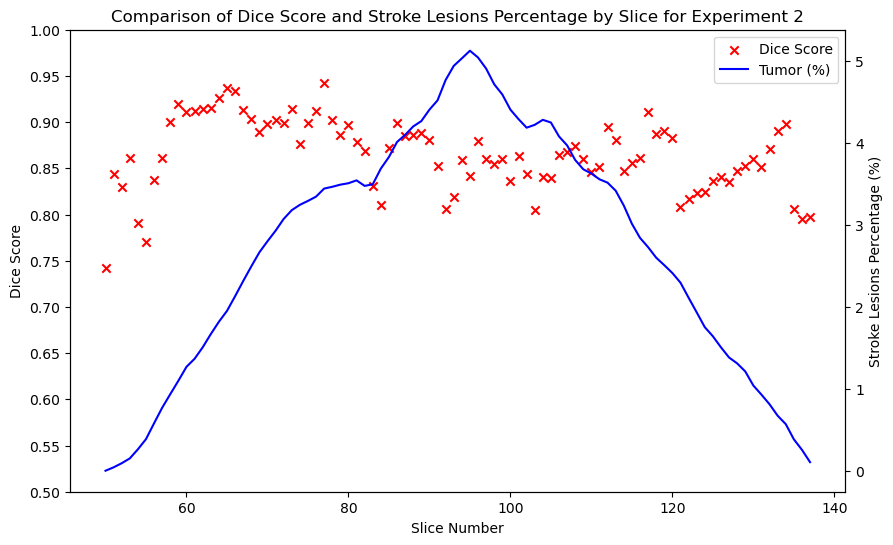}
\caption{Dice Score and stroke lesions Percentage per Slice in Experiment 2: This graph illustrates the relationship between the Dice score and the stroke lesions tissue percentage across individual MRI slices, as observed in the second experiment.}
\label{fig:exp2figure}
\end{figure}

\begin{table}[H]
\centering
\caption{Statistical Summary of Parameters for Experiment 2}
\label{tab:exp2data}
\begin{tabular}{@{}lccc@{}}
\toprule
 & \textbf{Fuzzy Threshold} & \textbf{Seeds} & \textbf{Dice Score} \\ 
\midrule
mean & 0.281 & 8.875 & 0.865 \\
std & 0.165 & 2.387 & 0.039 \\
min & 0.112 & 2.000 & 0.742 \\
max & 0.829 & 11.000 & 0.942 \\
\bottomrule
\end{tabular}
\end{table}

\subsection*{Experiment 3}
In the culmination of the optimization series, experiment three further refined the parameter tuning by incorporating the size of the dilation kernel used in the postprocessing stage into the optimization framework. This final iteration of experimentation sought to perfect the balance between the fuzzy threshold, the number of seeds, the Gaussian denoising standard deviation (\(\sigma\)), and the morphological operations that define the postprocessing contours of the segmented stroke lesions.

\begin{figure}[H]
\centering
\includegraphics[width=.487\textwidth]{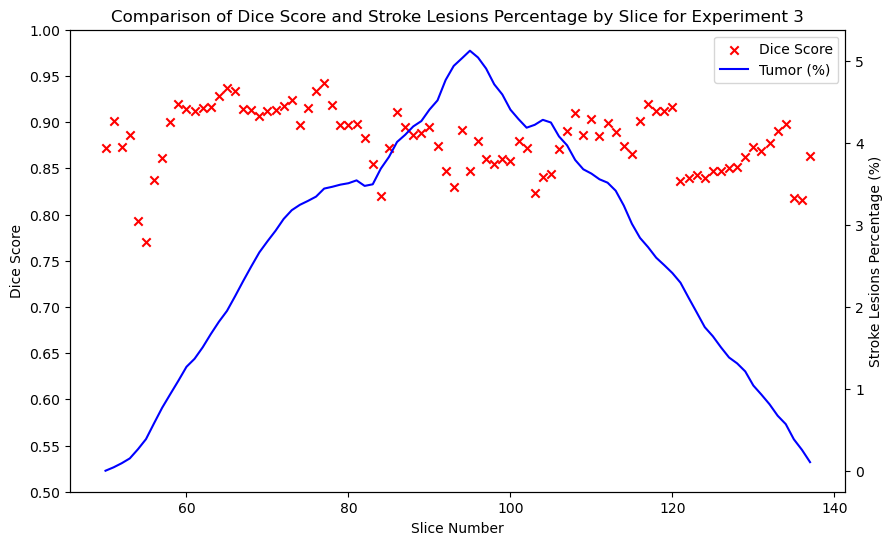}
\caption{Dice Score and stroke lesions Percentage per Slice in Experiment 3: Displayed here is the comparative graph of Dice scores against the percentage of stroke lesions tissue for each slice, as determined in the third experiment.}
\label{fig:exp3figure}
\end{figure}

\begin{table}[H]
\centering
\caption{Statistical Summary of Parameters for Experiment 2}
\label{tab:exp3data}
\begin{tabular}{@{}lccc@{}}
\toprule
 & \textbf{Fuzzy Threshold} & \textbf{Seeds} & \textbf{Dice Score} \\ 
\midrule
mean & 0.348 & 9.511 & 0.881 \\
std & 0.221 & 1.675 & 0.034 \\
min & 0.112 & 4.000 & 0.771 \\
max & 1.000 & 11.000 & 0.942 \\
\bottomrule
\end{tabular}
\end{table}

\subsection*{Computational Time}
In the final analysis of the optimization experiments for the FISRG algorithm, the computational time for each phase was a crucial metric. As the complexity of the parameter optimization increased across the experiments, there was a corresponding rise in the computational time required.

This increase in computational time is reflective of the iterative and exhaustive nature of the fine-tuning process, particularly as the number of parameters and the granularity of their adjustment scales up. The results provide valuable insights into the trade-offs between computational time and segmentation accuracy, with the detailed discussion of these dynamics reserved for the subsequent section.

\begin{table}[H]
\centering
\caption{Execution Time for each Experiment}
\label{tab:comptime}
\begin{tabular}{@{}rr@{}}
\toprule
Experiment & Computational Time \\
\midrule
1 & 73 min \\
2 & 148 min \\
3 & 503 min \\
\bottomrule
\end{tabular}
\end{table}

\begin{figure*}[b]
\centering
\includegraphics[width=1\textwidth]{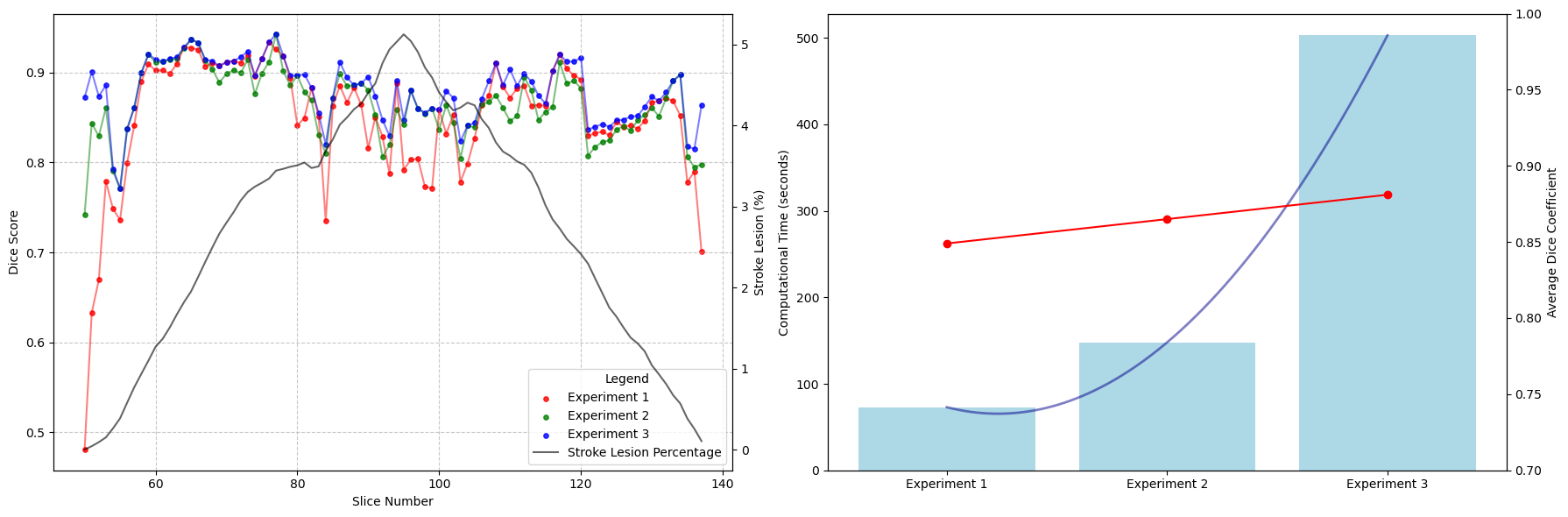}
\caption{Combined Analysis of Segmentation Accuracy and Computational Time: The combined plot juxtaposes the Dice scores from Experiments 1, 2, and 3 against each MRI slice number, with the stroke lesions percentage overlaid, and contrasts these results with the computational time required for each experiment. This visual comparison underscores the relationship between segmentation performance and computational investment across the study.}
\label{fig:combineddiscussion}
\end{figure*}

\section{Discussion}

\subsection{Accuracy Comparison Across Experiments}
For a comprehensive understanding of the segmentation accuracy across the three experiments, readers are directed to Fig. \ref{fig:combineddiscussion}, which visually delineates the performance of each approach.

The experimental results demonstrate that Experiment 1, while establishing a foundational understanding of the segmentation process, was less effective, particularly in slices with a minimal presence of stroke lesions tissue. Its performance dipped more significantly than the subsequent experiments in slices that proved challenging for all tested methods. Experiment 2 improved upon the limitations of the first, delivering better segmentation in slices with lower stroke lesions percentages. It consistently outperformed Experiment 1 and achieved parity with Experiment 3 in certain instances, but it did not surpass the latter in any slice.

Experiment 3 emerged as the most robust, maintaining a steady Dice coefficient across slices that traditionally present difficulties in segmentation, such as those with few stroke lesions tissue. It proved to be the superior method, yielding the highest scores across all scans. This suggests that a four-parameter optimization—encompassing the fuzzy threshold, the number of seeds, the Gaussian denoising sigma, and the dilation kernel size—is most conducive for the algorithm to perform optimally. Interestingly, all experiments exhibited similar trends in Dice score fluctuations, indicating that there are inherent challenges in the MRI scans that consistently affect segmentation performance, irrespective of the experiment.

\subsection{Balancing Accuracy and Computational Time}

In the evaluation of computational efficiency, Fig. \ref{fig:combineddiscussion} illustrates a crucial aspect of algorithm performance: the relationship between computational time and segmentation accuracy.

As shown in the referenced figure, there is a pronounced exponential increase in computational time across the experiments, while the improvement in the Dice coefficient accuracy exhibits a relatively modest linear progression. This pattern underscores a significant trade-off where gains in accuracy are obtained at the cost of substantially increased computational resources.

Although Experiment 3 achieved the highest Dice scores, suggesting superior segmentation capabilities, the computational demand associated with its extensive parameter fine-tuning raises practical concerns. The algorithm's exhaustive computational requirements may pose challenges, especially in scenarios where rapid processing is essential.

Experiment 2 presents itself as a balanced compromise. It delivers improved accuracy over Experiment 1 without incurring a proportionally high increase in computational time. This efficiency makes the configuration used in Experiment 2 an attractive option for the segmentation tasks addressed in this project, providing a pragmatic balance between accuracy and computational expenditure. Therefore, when considering both the accuracy and the computational aspects, Experiment 2 may represent the most viable approach for the segmentation methodology developed in this research.

\subsection{Algorithm Robustness and Limitations}

Analyzing the robustness of the algorithm as evidenced by the segmentation performance across varying stroke lesions percentages per slice in Fig. \ref{fig:combineddiscussion}, we can discern distinct patterns in the algorithm's behavior.

At the extremities of the MRI scan sequence—particularly the first and last scans where the stroke lesions presence is minimal—the segmentation accuracy of all experiments tends to decrease. This expected outcome can be attributed to the challenge of discerning extremely small stroke lesions regions from the surrounding tissue, especially after the Gaussian denoising step, which may inadvertently smooth these regions into near invisibility. It is observed that Experiments 2 and 3, which include fine-tuning of the Gaussian denoising sigma and the dilation kernel size, demonstrate improved segmentation in these challenging scans. The ability to adjust these parameters allows for better control over the intensity variations in small stroke lesions regions, resulting in higher Dice scores.

A notable fluctuation in performance is observed in the mid-sequence slices, approximately between slices 82 and 108, where the size of the stroke lesions region is at its peak. During this phase, all three experiments exhibit a decrease in segmentation stability. A plausible explanation for this phenomenon is the stroke lesions's expansion into areas of the brain characterized by low-intensity signals in T1-weighted MRI images, akin to those of the cerebrospinal fluid found in the subarachnoid space and brain ventricles. Given that the algorithm relies on intensity values to assign pixel region membership, the similarities in low-intensity values between stroke lesions tissue and cerebrospinal fluid-containing structures could lead to misclassification. Consequently, non-stroke lesionsous anatomical structures may be incorrectly segmented as part of the stroke lesions, adversely affecting the Dice coefficient.

The exploration of the algorithm's segmentation capability in relation to the varying stroke lesions percentages provides an opportunity to validate the proposed theory regarding its performance. As the stroke lesions percentage increases, particularly in the mid-sequence slices, a decrease in the Dice coefficient is observed, suggesting potential misclassification issues within these regions.

To investigate this theory, specific slices where the Dice coefficient notably diminishes were examined in detail. Slice 84, depicted in Fig. 9, serves as a prime example, where the segmentation performed by the algorithm in Experiment 1 erroneously extends into the subarachnoid layer. In Experiments 2 and 3, the misclassification appears to involve the brain's first or second ventricle, indicating that the segmentation challenge persists across experiments, albeit with varying degrees of severity.

Slice 103, illustrated in Fig. 10, presents a similar scenario. The segmentation in Experiment 1 inaccurately invades regions associated with the brain's ventricles. The modifications made in Experiments 2 and 3, which involve controlled blurring through fine-tuned denoising parameters, lead to improved results, thereby supporting the initial hypothesis. These observations confirm that while the algorithm demonstrates a strong capacity for accurate segmentation, its intensity-based classification strategy is prone to errors when encountering anatomical structures with similar intensity profiles to the stroke lesions tissue.

\begin{figure}[H]
\centering
\includegraphics[width=.487\textwidth]{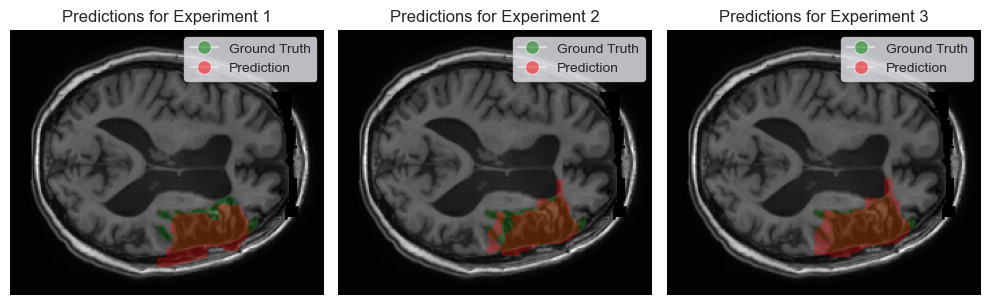}
\caption{ Segmentation Predictions for Slice 84 Across Experiments: This image displays the segmentation outcomes for slice 84 from each of the three experiments. The overlaid regions indicate the ground truth (in green) and the algorithm's predictions (in red), providing a visual comparison of the segmentation performance and the extent of overlap with the true stroke lesions boundaries.}
\label{fig:84}
\vspace{20pt} 
\includegraphics[width=.487\textwidth]{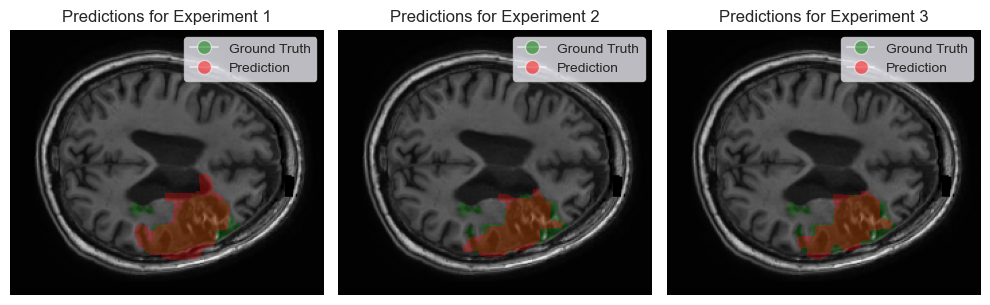}
\caption{Segmentation Predictions for Slice 103 Across Experiments: The image showcases the segmentation results for slice 103, highlighting the differences in the algorithm's predictions (in red) against the ground truth (in green) for each experiment. This comparison illustrates how the segmentation accuracy varies between experiments for a slice with a substantial stroke lesions presence.}
\label{fig:103}
\end{figure}
\vspace{-7pt} 
As the dataset progresses beyond the central brain slices where the stroke lesions reaches its maximum size, the Dice coefficient begins to stabilize once more across all experiments. This suggests that as the stroke lesions becomes more distinct from the surrounding brain anatomy, the segmentation algorithm's performance improves. However, this trend is disrupted by a notable decrease in the Dice coefficient between slices 120 and 121.

The sudden drop in the Dice coefficient at this juncture could be indicative of a significant change in the stroke lesions's structure, challenging the algorithm's ability to accurately segment the stroke lesions tissue. The gradual improvement in Dice scores following this drop supports the hypothesis that the algorithm is capable of adapting to changes in the stroke lesions's morphology over successive slices. The linear recovery in segmentation accuracy implies that, despite the initial impact of the structural change, the algorithm's underlying mechanics allow for recalibration to the new stroke lesions configuration.
\newpage
\begin{figure}[H]
\centering
\includegraphics[width=.487\textwidth]{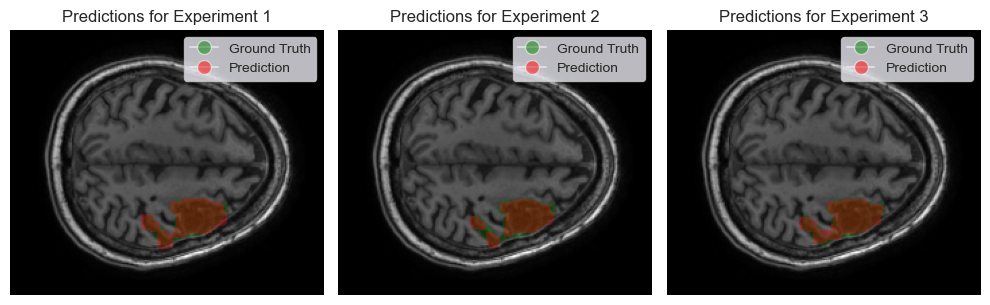}
\caption{Segmentation Predictions for Slice 120 Across Experiments: This figure shows the comparative segmentation predictions for slice 120, highlighting the regions identified by the algorithm (in red) against the ground truth (in green) for each experiment, demonstrating the performance before the observed drop in the Dice coefficient.}
\label{fig:120}
\end{figure}

\begin{figure}[H]
\centering
\includegraphics[width=.487\textwidth]{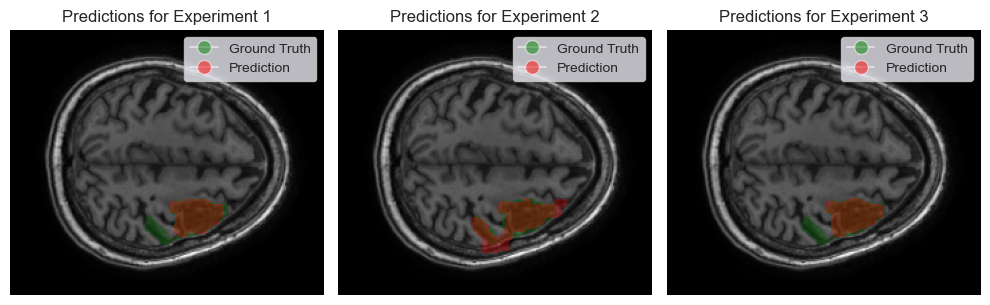}
\caption{Segmentation Predictions for Slice 121 Across Experiments: The figure presents the segmentation outcomes for slice 121, providing a visual account of how the algorithm's predictions (in red) compare to the ground truth (in green) immediately after the notable decrease in the Dice coefficient across experiments.}
\label{fig:121}
\end{figure}

The segmentation challenges presented in slices 120 and 121, as visualized in Figs. \ref{fig:120} and \ref{fig:121}, underscore the algorithm's sensitivity to the structural nuances of stroke lesions regions. The depicted slices illustrate a critical transition point where a slender bridge of stroke lesions tissue, previously connecting two larger stroke lesions masses, becomes markedly reduced in thickness. This alteration in the stroke lesions's morphology presents a significant challenge for the algorithm, which struggles to maintain the continuity of the stroke lesions region when faced with such minute connections.

In the case of Experiment 2, while the algorithm manages to accurately segment one of the stroke lesions regions, it erroneously compensates for the loss of connectivity by misclassifying adjacent non-stroke lesions tissue as part of the stroke lesions. This misclassification, particularly involving the subarachnoid space, reflects the algorithm's reliance on intensity values, which may not provide sufficient discriminative power in scenarios where the stroke lesions's structural integrity is compromised by such narrow linkages.

\subsection*{Robust points}

\begin{figure}[H]
\centering
\includegraphics[width=.487\textwidth]{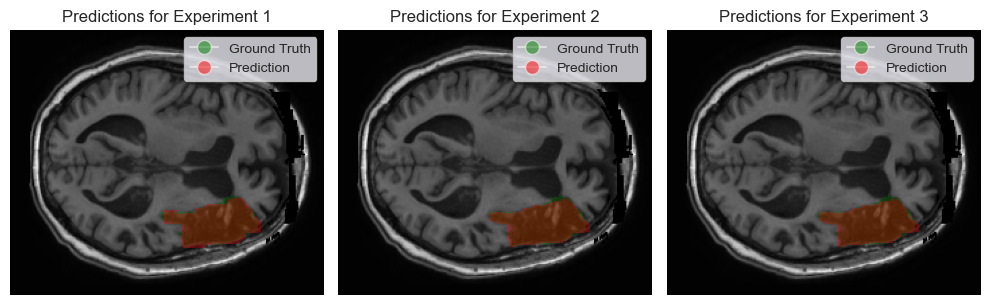}
\caption{Segmentation Performance on Slice 77 Across Experiments: This figure displays the segmentation results for slice 77, where the algorithm's predictions (in red) are shown alongside the ground truth (in green). Despite the stroke lesions's varied textures and intensities, the segmentation closely aligns with the ground truth across all three experiments, demonstrating the algorithm's capability to accurately delineate the stroke lesions boundaries.}
\label{fig:77}
\end{figure}

The algorithm shows a strong robustness against the \textit{complex textures within brain stroke lesions}. Despite the heterogeneity commonly found in stroke lesions regions, the algorithm is generally capable of maintaining a high segmentation accuracy. This behaviour is depicted in Fig. \ref{fig:77}.
When the characteristics of the stroke lesions remain \textit{consistent across slices}, the algorithm demonstrates excellent performance, achieving Dice coefficients in the range of 84\%-91\%. This indicates that the algorithm is highly effective in segmenting stroke lesions with stable features.

\subsection*{Limitations}

The algorithm occasionally struggles with \textit{abrupt changes in the stroke lesions's topology}, such as when a thin region of the stroke lesions, critical for the connection of two larger stroke lesions parts, becomes too narrow. This results in difficulty segmenting the disconnected stroke lesions region, potentially leading to a lower Dice coefficient.
Another limitation is the algorithm's tendency to \textit{misclassify anatomical structures with similar intensity to that of stroke lesion tissue}, such as the subarachnoid space or brain ventricles, as being part of the stroke lesion.

\section{Conclusion}

The performance of the FISRG algorithm, as explored in this research, has shed light on the distinct challenges and capabilities inherent in intensity-based segmentation methods, particularly when applied to MRI images of brain stroke lesions. This task is notably more complex than traditional tumor segmentation, primarily because stroke lesions often exhibit irregular boundaries and heterogeneous textures, unlike the more homogenous texture of common tumors like meningiomas, which tend to have smoother areas.

Despite these challenges, the algorithm has demonstrated a commendable robustness in managing the intricate and varied textures characteristic of stroke lesions, exceeding initial expectations in some aspects. Its success in accurately segmenting regions affected by stroke, despite the diverse intensities and unpredictable patterns within these lesions, is particularly noteworthy. This achievement is all the more impressive considering the algorithm's reliance solely on pixel intensities for segmentation, without the aid of additional textural or shape-based features. Such efficiency speaks to the sophisticated application of fuzzy logic and region growing techniques within the algorithm.

The algorithm's consistent efficacy across various sizes and morphologies of stroke lesions further underscores its clinical applicability. Although the optimization of parameters introduced a trade-off between computational time and accuracy, the balance struck in Experiment 2 – which combined higher average accuracy with reasonable computational demands – suggests a pragmatic approach for real-world medical applications.

In summary, this research not only confirms the effectiveness of the FISRG algorithm in segmenting brain stroke lesions from MRI scans but also opens avenues for further refinement and development. The challenges in distinguishing stroke lesions from non-lesioned regions with similar intensity profiles and adapting to sudden changes in lesion structure highlight areas for future improvement. The potential incorporation of spatial or contextual information into the algorithm could lead to enhanced precision and clinical relevance. Thus, this study lays a solid foundation for continuous advancement in this field, emphasizing the crucial role of image processing algorithms in aiding medical diagnostics and treatment strategies.

\section{Future Research}
\section*{Future Research}

Looking ahead, there are several promising avenues for advancing the capabilities of the segmentation algorithm detailed in this research. A detailed exploration of potential improvements and alternative approaches to the current algorithm is provided in Appendix A. These ideas offer a groundwork for future enhancements and adaptations of the segmentation methodology.

One particularly promising direction involves the application of Convolutional Neural Networks (CNNs) to the segmentation of stroke lesions. CNNs have revolutionized the field of image recognition and analysis due to their ability to learn complex patterns and features from vast amounts of data. Given the comprehensive dataset provided by the ATLAS project, there is a substantial opportunity to develop a CNN-based approach for stroke lesion segmentation. The depth and variety of the ATLAS dataset make it an ideal resource for training such models, providing a diverse range of lesion types and brain imaging scenarios. Recent studies, such as the work by Huo et al. (2022) on "MAPPING: Model Average with Post-processing for Stroke Lesion Segmentation" using the U-Net architecture, have already begun exploring these possibilities, demonstrating the potential of deep learning techniques in enhancing segmentation results \cite{huo2022mapping}.

By leveraging the strengths of deep learning, particularly in feature recognition and automatic pattern learning, a CNN-based model could potentially achieve greater accuracy and efficiency in segmenting stroke lesions compared to the current algorithm. The ability of CNNs to intuitively understand and process the intricate details within MRI scans may allow for superior differentiation between lesioned and non-lesioned brain tissues, even in cases with complex textures and irregular lesion shapes.

The prospect of integrating CNN methodologies with the existing segmentation techniques offers an exciting path forward. Such a hybrid approach could combine the nuanced understanding of lesion characteristics, as demonstrated by the current algorithm, with the advanced pattern recognition capabilities of CNNs. This could lead to more accurate, efficient, and clinically valuable segmentation tools, potentially outperforming existing methods.

\appendices
\section{Exploration of Alternative Approaches}
In the development of the FISRG algorithm for MRI brain stroke lesions segmentation, additional strategies were explored to enhance the Dice coefficient and refine the segmentation results. Two main approaches, while not integrated into the final methodology, provided significant insights and are worthy of discussion for potential application in future research.

\subsection{Refinement Using Chan Vese Active Contour Model}

One promising avenue investigated was the application of the Chan Vese active contour algorithm in the postprocessing stage to fine-tune the contours of the segmented stroke lesions. Inspired by the work mentioned in \cite{Yao2021SystematicMRI}, where the Chan Vese model was successfully applied following a Convolutional Neural Network (CNN) architecture to sharpen stroke lesions boundaries, this approach held potential for further refinement of the segmentation results. However, due to time constraints and the complexity associated with the fine-tuning required for the Chan Vese algorithm, this strategy was not pursued to completion within the scope of this study. Nonetheless, the initial promise of this technique suggests that, with adequate time and resources, it could significantly benefit future segmentation efforts.

\subsection{Leveraging Data Across Consecutive Slices}

Another approach considered was to incorporate information from adjacent MRI slices into the FISRG algorithm. This was envisioned to operate in two facets: seed selection and histogram matching. In seed selection, seeds from a previous slice with a high Dice coefficient would inform the selection in subsequent slices, leveraging spatial continuity. For histogram matching, the histogram of a representative stroke lesions slice would be used as a reference to standardize the intensity distribution across other slices, assuming minimal variation in stroke lesions texture. Despite the theoretical appeal given the consistent pixel intensity distribution in T1-weighted MRI images, this method did not yield an improvement in the overall Dice coefficient. This was likely due to the varied nature of stroke lesions textures between slices, which proved to be a significant variable affecting segmentation consistency.

These approaches, though not ultimately incorporated into the final workflow, underscore the iterative and exploratory nature of algorithm development in medical image analysis. Further studies with a focus on these methods, particularly with extended timeframes and enhanced computational resources, may uncover their potential to augment the accuracy and precision of stroke lesions segmentation in MRI images.

\section{Combined Statistical Summary of Dice Scores Across Experiments}
\begin{table}[H]
\centering
\label{tab:combinedDiceData}
\begin{tabular}{@{}lccc@{}}
\toprule
Statistic & Experiment 1 & Experiment 2 & Experiment 3 \\ 
\midrule
mean & 0.849 & 0.865 & 0.881 \\
std  & 0.071 & 0.039 & 0.034 \\
min  & 0.481 & 0.742 & 0.771 \\
max  & 0.934 & 0.942 & 0.942 \\
\bottomrule
\end{tabular}
\end{table}

\section{Open Source Code}
All the code developed for this project, encompassing the implementation of the FISRG algorithm and the subsequent analysis, is open source and publicly available. This initiative aligns with our commitment to promoting transparency, collaboration, and the advancement of scientific knowledge in the field of medical image processing. For detailed insight into the codebase, including documentation and usage instructions, you are encouraged to visit our GitHub repository. The repository can be accessed at \href{https://github.com/Mawio02/FISRG-for-Automated-Lesion-After-Stroke-Segmentation-in-MRI}{GitHub Project}.

\section*{Acknowledgment}
I would like to express my deepest gratitude to Dr. Enrique Nava Baro for his invaluable guidance and mentorship throughout this project. Dr. Nava Baro was instrumental in shaping the direction of this research, from proposing the initial topic to meticulously reviewing its progression. His insightful feedback, encouragement to explore various approaches, and unwavering support were pivotal in navigating the challenges encountered during this study. His expertise and dedication have been a source of inspiration and have greatly contributed to the success of this project.

I am also profoundly thankful to Dr. Ezequiel Lopez Rubio for his expert advice in the creation of the visual and graphical elements of this research. 

Their combined expertise and support have been invaluable, and I am deeply appreciative of their contributions to my research journey.

\bibliographystyle{IEEEtran}
\bibliography{references}

\end{document}